\documentclass{llncs}
\usepackage{etoolbox}
\newtoggle{proofs}
\toggletrue{proofs}   % only in llncs style
\usepackage[version=0.96]{pgf}
\usepackage{tikz}
\usetikzlibrary{arrows,shapes,snakes,automata,backgrounds,petri}
\usepackage{stmaryrd,epsfig,color,moreverb}
\usepackage{graphicx}

\usepackage{mathtools} 
\usepackage{amsmath}
\usepackage{amssymb}
\usepackage{environ, thmtools, thm-restate}  % Advanced theorem handling
\usepackage{enumerate}

\usepackage{eucal}
\usepackage{wasysym}

\usepackage{cases}

\NewEnviron{reslemma}[2][]
  {\begin{restatable}[#1]{mylem}{#2}%
    \label{lem:#2}
    \BODY
   \end{restatable}%
  }

\iftoggle{proofs}{%
\declaretheorem[name=Notation,numberlike=definition]{notation}
\declaretheorem[name=Theorem,numberlike=definition]{mythm}
\declaretheorem[name=Lemma,numberlike=definition]{mylem}
\declaretheorem[name=Proposition,numberlike=definition]{myproposition}
\declaretheorem[name=Theorem,numberwithin=section]{appthm}
\declaretheorem[name=Lemma,numberwithin=section]{applem}
\declaretheorem[name=Definition,numberwithin=section]{appdef}
}
{%
\declaretheorem[name=Theorem,numberwithin=section]{theorem}

\declaretheorem[name=Lemma,numberlike=theorem]{lemma}
\declaretheorem[name=Definition,numberlike=theorem]{definition}
\declaretheorem[name=Example,numberlike=theorem]{example}
\declaretheorem[name=Theorem,numberlike=theorem]{mythm}
\declaretheorem[name=Lemma,numberlike=lemma]{mylem}

}

\newcommand{\atoms}{\mathcal{T}}
\newcommand{\ccomp}[0]{\mid}

\newcommand{\wif}[3]{\textrm{if } {#1} \;\textrm{then}\;{#2}\;\textrm{else}\;{#3}}

\newcommand{\lending}{{\circledcirc}}
\newcommand{\nolending}{{\ocircle}}

\newcommand{\urgent}[1][]{\mathcal{U}_{{#1}}}

\newcommand{\curgent}[2][]{\urgent[{#1}]^{#2}}

\newcommand{\trans}[1]{\ensuremath{\,[\/{#1}\/\rangle}\,}

\newcommand{\pre}[1]{\ensuremath{\!~^\bullet{#1}}}

\newcommand{\post}[1]{\ensuremath{{#1} {^\bullet}}}

\newcommand{\nat}{\ensuremath{\mathbb{N}}}

\newcommand{\flt}[1]{\ensuremath{[\![{#1}]\!]}}

\newcommand{\contrtopnl}{\ensuremath{\mathcal{P}}}

\newcommand{\Princ}{{\it Part}}
\newcommand{\aname}{\mathcal{A}}
\newcommand{\cname}{\mathcal{C}}

\newcommand{\princsym}{\pi}
\newcommand{\princ}[2][]{\princsym_{#1}({#2})}
\newcommand{\invprinc}[2][]{\princsym_{#1}^{-1}({#2})}

\newcommand{\payoffsym}[1][]{\Omega_{#1}}

\newcommand{\oksym}{{\it ok}}

\newcommand{\powset}[1]{\wp(#1)}

\newcommand{\seq}[1]{\langle {#1} \rangle}
\newcommand{\mkset}[1]{\overline{#1}}

\renewcommand{\epsilon}{\varepsilon}
\newcommand{\hidden}[1]{}

\newcommand{\pmv}[1]{\ensuremath{\mathsf{#1}}}

\newcommand{\atom}[1]{\textup{\textsf{#1}}}

\newcommand{\coimp}{\twoheadrightarrow}

\newcommand{\irule}[2]{
  \begin{array}{c}
    #1  \\ \hline
    #2
  \end{array}}

\newcommand{\imp}{\rightarrow}
\newcommand{\nrule}[1]{{\scriptsize \textsc{#1}}}
\newcommand{\sem}[1]{\mbox{\ensuremath{\llbracket{#1}\rrbracket}}}

\newcommand{\pcl}{\textup{PCL\;}}

\newcommand{\setenum}[1]{\{#1\}}
\newcommand{\setcomp}[2]{\{{#1} \;\mid\; {#2}\}}

\newcommand{\pname}{\mathcal{D}}

\title{Lending Petri nets and contracts}
\author{Massimo Bartoletti \and Tiziana Cimoli \and G.~Michele Pinna}
\institute{Dipartimento di Matematica e Informatica, Universit\`a degli Studi di Cagliari, Italy}

\begin{document}
\pagestyle{plain}

\maketitle

\begin{abstract}
Choreography-based approaches to service composition
typically assume that, after a set of services has been
found which correctly play the roles prescribed by
the choreography, each service respects his role.
Honest services are not protected against adversaries.
We propose a model for contracts based on an extension of Petri nets, 
which allows services to protect themselves while still
realizing the choreography.
We relate this model with Propositional Contract Logic,
by showing a translation of formulae into our Petri nets
which preserves the logical notion of agreement, 
and allows for compositional verification.
\end{abstract}

\section{Introduction}

Many of today's human activities, from business and financial
transactions, to collaborative and social applications, 
run over complex interorganizational systems, 
based on service-oriented computing (SOC) and cloud computing technologies. 
These technologies foster the implementation of complex software systems
through the composition of basic building blocks, called \emph{services}.
Ensuring reliable coordination of such components is fundamental to
avoid critical, possibly irreparable problems, ranging from economic
losses in case of commercial activities, to risks for human life in
case of safety-critical applications.

Ideally, in the SOC paradigm 
an application is constructed by dynamically discovering and composing services
published by different organizations.
Services have to \emph{cooperate} to achieve the overall goals,
while at the same time they have to \emph{compete}
to achieve the specific goals of their stakeholders.
These goals may be conflicting,
especially in case of mutually distrusted organizations.
Thus, services must play a double role:
while cooperating together, they have to protect themselves
against other service's misbehavior (either unintentional or malicious).

The lack of precise guarantees about the reliability and security of
services is a main deterrent for industries wishing to move
their applications and business to the cloud~\cite{Armbrust10cacm}.
Quoting from~\cite{Armbrust10cacm},
``absent radical improvements in security technology, we expect that
users will use contracts and courts, rather than clever security
engineering, to guard against provider malfeasance''.

Indeed, contracts are already a key ingredient in the design of SOC applications.
A \emph{choreography} is a specification of the overall behavior 
of an interorganizational process.
% choreography languages Web Services Description Language 
% (WS-CDL,~\cite{wscdl}), BPEL4Chor~\cite{bpel4chor}.
This \emph{global} view of the behavior is projected into 
a set of \emph{local} views, which specify the behavior expected from
each service involved in the whole process.
The local views can be interpreted as the service contracts:
if the actual implementation of each service respects its contract,
then the overall application must be guaranteed to behave correctly.

There are many proposals of formal models for contracts in the literature, which
we may roughly divide into ``physical'' and ``logical'' models.
Physical contracts take inspiration mainly from formalisms for concurrent
systems (e.g.\ Petri nets~\cite{Aalst10cj}, 
event structures~\cite{Hildebrandt10places,BCPZ12places}, and various sorts 
of process algebras~\cite{bhty10,Bravetti08tgc,Bravetti07fsen,Castagna09contracts,Honda08popl}), and
they allow to describe the interaction of services in terms
of response to events, message exchanges, \emph{etc.}
On the other side, 
logical contracts are typically expressed as formulae of suitable logics, 
which take inspiration and extend e.g.\ 
modal~\cite{Abadi93access,Garg08modal},
intuitionistic~\cite{Abadi93logical,BZ10lics}, linear~\cite{Abadi93logical},
deontic~\cite{Prisacariu11jlap} logics
to model high-level concepts such as promises, obligations, prohibitions,
authorizations, \emph{etc.}

Even though logical contracts are appealing,
since they aim to provide formal models and reasoning tools for real-world 
Service Level Agreements,
existing logical approaches have not had a great impact on the 
design of SOC applications.
A reason is that there is no evidence on how to relate high-level properties
of a contract with properties of the services which have to realize it.
The situation is  decidedly better in the realm of physical contracts,
where the gap between contracts and services is narrower.
Several papers,
e.g.~\cite{Bravetti08tgc,Bravetti07fsen,Bravetti07sc,Honda08popl,Aalst10cj}, 
address the issue of relating properties of a choreography
with properties of the services which implement it
(e.g.\ deadlock freedom, communication error freedom, session fidelity),
in some cases providing automatic tools to project the choreography 
to a set services which correctly implements it.

A common assumption of most of these approaches is that
services are \emph{honest}, i.e.\ their behavior always adheres 
to the local view.
For instance, if the local view takes the form of a behavioral type,
it is assumed that the service is typeable, and that its type
is a subtype of the local view.
Contracts are only used in the ``matchmaking'' phase:
once, for each local view projected from the choreography, 
a compliant service has been found, 
then all the contracts can be discarded.

We argue that the honesty assumption is not suitable in the case
of interorganizational processes, where services may pursue
their providers goals to the detriment to the other ones.
For instance, consider a choreography which prescribes that
a participant {\pmv A} performs action $a$
(modeling e.g.\ ``pay \$100 to {\pmv B}''),
and that {\pmv B} performs $b$ 
(e.g.\ ``provide {\pmv A} with 5GB disk storage'').
If both {\pmv A} and {\pmv B} are honest, 
then each one will perform its due action, 
so leading to a correct execution of the choreography.
However, since providers have full control of the services they run,
there is no authority which can force services to be honest.
So, a malicious provider can replace a service validated w.r.t.\ its contract, 
with another one:
e.g., {\pmv B} could wait until {\pmv A} has done $a$,
and then ``forget'' to do $b$.
Note that {\pmv B} may perform his scam
while not being liable for a contract violation,
since contracts have been discarded after validation.

In such competitive scenarios, the role of contracts is twofold.
On the one hand, they must guarantee that their composition
complies with the choreography:
hence, in contexts where services are honest,
the overall execution is correct.
On the other hand, contracts must protect services from malicious ones:
in the example above, the contract of {\pmv A} must ensure that,
if {\pmv A} performs $a$, then {\pmv B} will either do $b$, or
he will be considered culpable of a contract violation.

In this paper, we consider physical contracts modeled as Petri nets,
along the lines of~\cite{Aalst10cj}.
In our approach we can both
start from a choreography (modeled as a Petri net)
and then obtain the local views by projection,
as in~\cite{Aalst10cj},
or start from the local views, 
i.e.\ the contracts published by each participant,
to construct a choreography which satisfies the goals of everybody.
Intuitively, when this happens the contracts admit an \emph{agreement}.

A crucial observation of~\cite{BCZ13post} 
is that if contracts admit an agreement,
then some participant is not protected, and \emph{vice-versa}.
The archetypical example is the one outlined above.
Intuitively, if each participant waits until someone else has performed her action,
then everyone is protected,
but the contracts do not admit an agreement because of the deadlock.
Otherwise, if a participant does her action without waiting,
then the contracts admit an agreement, 
but the participant who makes the first step is not protected.
This is similar to the proof of impossibility of fair exchange
protocols without a trusted third party~\cite{EvenYacobi80}.

To overcome this problem, we introduce \emph{lending Petri nets} (in short, LPN).
Roughly, an LPN is a Petri net where some places may give tokens ``on credit''.
Technically, when a place gives a token on credit 
its marking will become negative.
This differs from standard Petri nets, where markings are always nonnegative.
The intuition is that if a participant takes a token on credit,
then she is obliged to honour it --- otherwise she is culpable of a contract 
violation.

Differently from the Petri nets used in~\cite{Aalst10cj},
LPNs allow for modeling contracts which, at the same time,
admit an agreement (more formally, \emph{weakly terminate}) 
and protect their participants.
% We define composition of LPNs, and we prove that it is the
% inverse of choreography projection of~\cite{Aalst10cj}.
LPNs preserve one of the main results of~\cite{Aalst10cj},
i.e.\ the possibility of proving that an application respects a choreography,
by only locally verifying the services which compose it.
More precisely, we project a choreography to a set of local views,
independently refine each of them, and be guaranteed
then the composition of all refinements respects the choreography.
This is stated formally in Theorem~\ref{th:weakly term and accordance}.

The other main contribution is a relation between 
the logical contracts of~\cite{BZ10lics} and LPN contracts.
More precisely, we consider contracts expressed in (a fragment of) 
Propositional Contract Logic (\pcl\!\!), 
and we compile them into LPNs.
Theorem~\ref{th:pnl conf are states} states that 
a \pcl contract admits an agreement if and only if its compilation weakly terminates.
Summing up, Theorem~\ref{th:lpn-refinement} states that
one can start from a choreography represented as a logical contract,
compile it to a physical one, and then use 
Theorem~\ref{th:weakly term and accordance} to project it 
to a set services which correctly implement it, and which are protected
against adversaries.
Finally, Theorem~\ref{th:enc-u} relates logical and physical characterizations
of \emph{urgent} actions, \emph{i.e.} those actions which must be
performed in a given state of the contract.

%
%\vspace{-5pt}
%
%\paragraph{Structure of the paper.}
%% The rest of this paper is organized as follows.
%In Sect.~\ref{sec:nets} we review Petri nets.
%We  introduce lending Petri nets in Sect.~\ref{sec:lendingnets},
%and in Sect.~\ref{sec:lnp-contracts} we use them in a model for contracts.
%In Sect.~\ref{sec:pcl} we recap the logic \pcl\!.
%In Sect.~\ref{sec:translation} we show how to construct a physical contract
%from a logical one, and we state our main results.
%Finally, in Sect.~\ref{sec:conclusions} we 
%% discuss some related work and 
%draw some conclusions.

\vspace{-5pt}
\section{Nets}\label{sec:nets}

We briefly review Petri nets~\cite{Reisig-Book} and the token game. 
We consider Petri nets labeled on a set $\mathcal{T}$, and
% (differently from the usual definition), 
(perhaps a bit unusually)
the labeling is also on places.

%\begin{definition} \label{de:petri net}
A \emph{labeled Petri net} is a 5-tuple $\langle
S, T, F, \Gamma, \Lambda\rangle$, where
%\begin{itemize}
%
%\item 
 $S$ is a set of {\em places}, and $T$ is a set of {\em transitions} (with $S \cap T = \emptyset$), 
%
%\item 
 $F \subseteq (S\times T) \cup (T\times S)$ is the \emph{flow relation}, 
%
%\item 
and $\Gamma : S \to \mathcal{T}$, $\Lambda : T \to \mathcal{T}$
are partial \emph{labeling function} for places and transitions, respectively.
%
%\end{itemize} 
%\end{definition} 
 %
 Ordinary (non labeled) Petri nets are those where the two labeling functions are always undefined (\emph{i.e.} equal to $\bot$).
 We require that for each  $t \in T$,  $F(t,s) > 0$ for some
 place $s \in S$, i.e.\ a transition cannot happen \emph{spontaneously}.
 Subscripts on the net name carry over the names of the net components.
 As usual, we define the \emph{pre-set} and \emph{post-set} of a transition/place: 
 $\pre{x} = \{y\in T\cup S\mid F(y,x)>0\}$ and 
 $\post{x} = \{y\in T\cup S\mid F(x,y)>0\}$, respectively. 
 These are extended to subsets of transitions/places in the obvious way. 
 
 A \emph{marking} is a function $m$ from places to natural numbers (\emph{i.e.} a multiset over places), 
 which represents the state of the system modeled by the net.
 %
% \begin{definition}
%   \label{de:marked petri net}
   A \emph{marked} Petri net is a pair $N = (\langle
   S, T, F, \Gamma, \Lambda\rangle, m_0)$, where
%    \begin{itemize}
%    \item 
    $\langle S, T, F, \Gamma, \Lambda\rangle$ is a labelled Petri net, and 
%    \item 
    $m_0 : S \to \nat$ is the {\em initial marking}.
%   \end{itemize} 
% \end{definition} 
 
 The dynamic of a net is described by the execution of transitions at markings.
 Let $N$ be a marked net
(hereafter we will just call net a marked net).
% unless it is strictly needed to specify otherwise). 
 %
 A transition $t$ is enabled
 at a marking $m$ if the places in the pre-set of $t$ contains enough tokens 
 (\emph{i.e.} if $m$ \emph{contains} the pre-set of $t$).
 Formally,  $t \in T$ is {\em enabled} at $m$ if
 $m(s)\geq F(s,t)$ for all $s\in \pre{t}$.
 In this case, to indicate that the execution of $t$ in $m$ produces
 the new marking $m'(s) = m(s) - F(s,t) + F(t,s)$, we write $m \trans{t} m'$, and we call it a 
 \emph{step}\footnote{The word step is usually reserved to the execution of a subset of transitions, but
 here we prefer to stress the computational interpretation.}.
 This notion is lifted, as usual, to multisets of transitions.
 
 The notion of step leads to that of \emph{execution} of a net. 
 Let $N = (\langle S, T, F, \Gamma, \Lambda\rangle,$ $m_0)$ be a net, and let $m$ be a marking.
The \emph{firing sequences} starting at $m$ are defined as follows:
% \begin{itemize}
%  \item 
(a)  $m$ is a firing sequence, and 
%  \item 
%  
(b)  if $m\trans{t_1}m_1 \cdots m_{n-1}\trans{t_n}m_n$ is a firing sequence and $m_{n}\trans{t}m'$ is a 
        step, then 
        $m\trans{t_1}m_1 \cdots m_{n-1}\trans{t_n}m_n\trans{t}m'$ is a firing sequence.
% \end{itemize}
 %
 A marking $m$ is \emph{reachable} iff there exists a firing sequence starting at $m_0$ leading to it. 
 The set of reachable markings of a net $N$ is denoted with $\mathsf{M}(N)$. 
%
%\begin{definition}\label{de:safenets} 
 A net $N = (\langle S, T, F, \Gamma, \Lambda\rangle, m_0)$ is \emph{safe} when
 each marking $m\in \mathsf{M}(N)$ is such 
 that $m(s) \leq 1$ for all $s\in S$. 
%\end{definition}
 
 A \emph{trace} can be associated to each firing sequence, which is the word on $\mathcal{T}^{*}$
 obtained by the firing sequence considering just the (labels of the) transitions and forgetting 
 the markings: 
 if $m_0\trans{t_1}m_1 \cdots m_{n-1}\trans{t_n}m_n$ is a firing sequence of $N$, the associated
 trace is $\Lambda(t_1t_2\dots t_n)$. The trace associated to $m_0$ is the empty word $\epsilon$.
 If the label of a transition is undefined then the associated word is the empty one. 
 The traces of a net $N$ are denoted with $\mathit{Traces}(N)$.

%\paragraph{Subnets.}
A \emph{subnet} is a net obtained 
by restricting places and transitions of a net, and 
correspondingly the flow relation and the initial marking.
% 
% \begin{definition}
%   \label{de:subnet}
   Let $N = (\langle S, T, F, \Gamma, \Lambda\rangle,$ $m_0)$ be a net, and let $T'\subseteq T$.
   We define the subnet generated by $T'$ as the net
   $N|_{T'} = (\langle S', T', F', \Gamma', \Lambda'\rangle,  m_0')$, where
%     \begin{itemize}
%     \item
        $S' = \{s\in S\ |\ F(t,s)>0$ or $F(s,t)>0$ for $t\in T'\}\cup
         \{s\in S\ |\ m_0(s)>0\}$,
%     \item
        $F'$ is the flow relation restricted to $S'$ and $T'$,
%     \item
        $\Gamma'$ is obtained by $\Gamma$ restricting to places in $S'$,
%     \item
        $\Lambda'$ is obtained by $\Lambda$ restricting to transitions in $T'$, and
%     \item
        $m_0'$ is obtained by $m_0$ restricting to places in $S'$   
%    \end{itemize}
% \end{definition}

A net property (intuitively, a property of the system modeled as a Petri net) can be characterized 
in several ways, \emph{e.g.} as a set of markings (states of the system). 
The following captures the intuition that, notwithstanding the state (marking) reached by the 
system, it is always possible to reach a state satisfying the property. 
%\begin{definition}
% [Weakly terminating nets] 
A net $N$ \emph{weakly terminates in} a set of  markings $\mathcal{M}$
iff
$\forall m\in \mathsf{M}(N)$, 
there is a firing sequence starting at $m$ and leading 
to a marking in $\mathcal{M}$. 
Hereafter, we shall sometimes say that $N$ weakly terminates 
(without referring to any $\mathcal{M}$) when
the property is not relevant or clear from the context.
%\end{definition}

% \paragraph{Occurrence Nets.}
We now introduce occurrence nets. 
The intuition behind this notion is the following: 
regardless how tokens are produced or consumed,
an occurrence net guarantees that each transition can occur only once 
(hence the reason for calling them occurrence nets). 
We adopt the notion proposed by van Glabbeek and Plotkin in \cite{GP:CS},
namely 1-occurrence nets. 
For a multiset $M$, we denote by $\flt{M}$ the multiset defined as 
$\flt{M}(a) = 1$ if $M(a) > 0$ and $\flt{M}(a) = 0$ otherwise.
%
% \begin{definition}
% \label{def:state of a net}
   A {\em state} of a net 
   $N = (\langle S, T, F, \Gamma, \Lambda\rangle, m_0)$ 
   is any finite multiset $X$ of $T$ such that the function 
   $m_X : S \rightarrow \mathbb{Z}$ given by 
   $m_X(s) = m_0(s) + \sum_{t\in T}X(t)\cdot (F(t,s) - F(s,t))$, 
   for all $s\in S$, 
   is a reachable marking of the net. 
   We denote by $\mathsf{St}(N)$ the states of $N$.
% \end{definition}
%
 A state contains (in no order) all the occurrence of the transitions that have been fired to reach
 a marking. 
 Observe that a trace of a net is a suitable linearization of the elements of a state $X$.
We use the notion of state to formalize occurrence nets.
% 
% \begin{definition}
% \label{de:occurrence net}
An \emph{occurrence net} 
$O = (\langle S, T, F, \Gamma, \Lambda\rangle, m_0)$ is a net where
each state is a set, i.e. $\forall X\in \mathsf{St}(N).\; X = \flt{X}$.
% \end{definition}
 
% \paragraph{Correctly labeled nets.}
 
% We end this section stating when a net is \emph{correctly} labeled.
% For a net to be correctly labeled it is required that 
% all the transitions in the pre-set of a labeled place 
% have the same label. 
%
%\begin{definition}\label{de:corr-labeled net} 
A net is \emph{correctly labeled} iff 
% $\forall s, t, t' : \; t,t' \in\pre{s}\ \implies \Lambda(t) = \Lambda(t')$.
% $\forall s. \forall t, t' \in\pre{s}.\; \Lambda(t) = \Lambda(t')$.
$\forall s. \forall t, t' \in\pre{s}.\; \Gamma(s) \neq \bot \!\implies\! \Lambda(t) = \Lambda(t') = \Gamma(s)$.
%\end{definition}
%
Intuitively, this requires that all the transitions putting a token 
in a labeled place represent the same action.

\section{Nets with lending places}\label{sec:lendingnets}

We now relax the conditions under which transitions may be executed,
by allowing a transition to consume tokens from a place $s$ 
even if the $s$ does not contain enough tokens.
Consequently, we allow markings with negative numbers. 
When the number of tokens associated to a place becomes negative, we say that
they have been done \emph{on credit}.
We do not permit this to happen in all places,
but only in the \emph{lending} places (a subset $\mathcal{L}$ of $S$).
Lending places are depicted with a double circle.

\begin{definition}
\label{de:lendnet}
A \emph{lending} Petri net (LPN) is a triple 
$(\langle S, T, F, \Gamma, \Lambda\rangle, m_0, \mathcal{L})$ where
$(\langle S, T, F, \Gamma, \Lambda\rangle,m_0)$ is a marked Petri net,
and $\mathcal{L}\subseteq S$ is the set of \emph{lending} places. 
\end{definition}
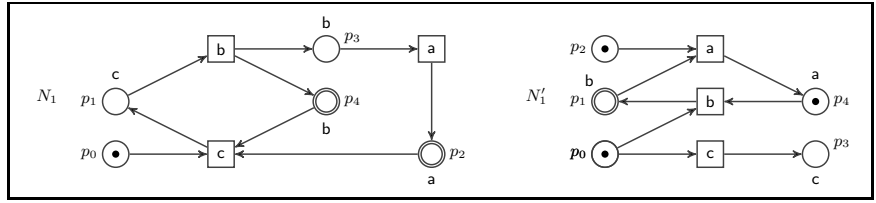
\begin{figure}[t!]
\begin{center}
\fbox{
%\hspace{20pt}
\scalebox{0.7}{
{\begin{tikzpicture}[>=stealth'] 
 \tikzstyle{lplace}=[circle,thick,draw=black!75,minimum size=4mm]
 \tikzstyle{place}=[circle,thick,draw=black!75,minimum size=5mm]
 \tikzstyle{transition}=[rectangle,thick,draw=black!75,minimum size=5mm]
 \tikzstyle{edge}=[->,thick,draw=black!75]
 \tikzstyle{edgered}=[->,thick,draw=red!75]
 \tikzstyle{edgeblu}=[->,thick,draw=blue!75]
 \path ( 0, 1) node [place] (p1) {}
       ( 0, 1) node [xshift=-5mm] {$p_1$}
       ( 0, 1) node [yshift=5mm] {\atom c}
       ( 2, 0) node [transition] (c) {\atom{c}}
       ( 2, 2) node [transition] (b) {\atom{b}} 
       ( 6, 2) node [transition] (a) {\atom{a}}
       ( 6, 0) node [place] (p2) {}
       ( 6, 0) node [xshift=5mm] {$p_2$}
       ( 6, 0) node [yshift=-5mm] {\atom a}
       ( 4, 2) node [place] (p3) {}
       ( 4, 2) node [xshift=5mm,yshift=2mm] {$p_3$}
       ( 4, 2) node [yshift=5mm] {\atom b}
       ( 4, 1) node [place] (p4) {}
       ( 0, 0) node [place] (p0) {}
       ( 0, 0) node [xshift=-5mm] {$p_0$}
       ( 4, 1) node [xshift=5mm] {$p_4$}
       ( 4, 1) node [yshift=-5mm] {\atom b}
       ( 4, 1) node [lplace] {}
       ( 6, 0) node [lplace] {}
       ( 0, 0) node [token] {};
       \node at (-1.3, 1.1) {$N_1$};
       \draw [edge] (c) to (p1);
       \draw [edge] (b) to (p4);
       \draw [edge] (b) to (p3);
       \draw [edge] (p1) to (b);
       \draw [edge] (a) to (p2);
       \draw [edge] (p2) to (c);
       \draw [edge] (p3) to (a);
       \draw [edge] (p4) to (c);
       \draw [edge] (p0) to (c);
\end{tikzpicture}}
\hspace{20pt}
\begin{tikzpicture}[>=stealth'] 
 \tikzstyle{lplace}=[circle,thick,draw=black!75,minimum size=4mm]
 \tikzstyle{place}=[circle,thick,draw=black!75,minimum size=5mm]
 \tikzstyle{transition}=[rectangle,thick,draw=black!75,minimum size=5mm]
 \tikzstyle{edge}=[->,thick,draw=black!75]
 \tikzstyle{edgered}=[->,thick,draw=red!75]
 \tikzstyle{edgeblu}=[->,thick,draw=blue!75]
 \path ( 0, 0) node [place] (p0) {}
       ( 0, 0) node [xshift=-5mm] {$p_0$}
       ( 0, 0) node [token] {}
       ( 0, 1) node [place] (p1) {}
       ( 0, 1) node [xshift=-5mm] {$p_1$}
       ( 0, 1) node [xshift=-3mm,yshift=4mm] {\atom b}
       ( 0, 1) node [lplace] {}
       ( 0, 2) node [place] (p2) {}
       ( 0, 2) node [xshift=-5mm] {$p_2$}
       ( 0, 2) node [token] {}
       ( 2, 1) node [transition] (b) {\atom{b}}
       ( 2, 0) node [transition] (c) {\atom{c}}
       ( 2, 2) node [transition] (a) {\atom{a}} 
       ( 4, 0) node [place] (p3) {}
       ( 4, 0) node [xshift=5mm,yshift=2mm] {$p_3$}
       ( 4, 0) node [yshift=-5mm] {\atom c}
       ( 0, 0) node [place] (p0) {}
       ( 4, 1) node [place] (p4) {}
       ( 4, 1) node [xshift=5mm] {$p_4$}
       ( 4, 1) node [yshift=5mm] {\atom a}
       ( 4, 1) node [token] {}
       ( 0, 0) node [xshift=-5mm] {$p_0$};
       \node at (-1.3, 1.1) {$N_1'$};
       \draw [edge] (p0) to (b);
       \draw [edge] (p1) to (a);
       \draw [edge] (p2) to (a);
       \draw [edge] (p4) to (b);
       \draw [edge] (p0) to (c);
       \draw [edge] (b) to (p1);
       \draw [edge] (c) to (p3);
       \draw [edge] (a) to (p4);
\end{tikzpicture}}
}
\end{center}
\vspace{-15pt}
\caption{Two lending Petri nets.}
\label{fig:carl net}
\end{figure}
\begin{example}\label{ex:carl net}
Consider the LPN $N_1$ in Fig.~\ref{fig:carl net}.
%
%\noindent
The places $p_2$ and $p_4$ are lending places. 
The set of labels of the transitions 
 is $\mathcal{T} = \setenum{\atom{a}, \atom{b}, \atom{c}}$, 
and the set of labels of the places is
$\mathcal{G} = \mathcal{T}$.
The labeling is 
$\Gamma(p_1) = \atom{c}, \Gamma(p_2) = \atom{a}$ and 
$\Gamma(p_4) = \Gamma(p_3) = \atom{b}$ (the place $p_0$ is unlabeled). 
% Clearly the net is correctly labeled.
\end{example}
 
The notion of step is adapted to take into account this new kind of places. 
Let $N$ be an LPN, let $t$ be a transition in $T$, and let $m$ be a marking. 
We say that $t$ is \emph{enabled} at $m$ iff 
$\forall s\in \pre{t}.\; m(s) \leq 0 \implies s\in\mathcal{L}$. 
The evolution of $N$ is defined as before, with the difference that 
the obtained marking is now a function from places to $\mathbb{Z}$ (instead of~$\nat$).
This notion matches the intuition behind of lending places: 
we allow a transition to be
executed even when some of the transitions that are a pre-requisite have not been executed yet. 
%
%This leads us to the following definition.
%
\begin{definition}
\label{de:correct-mark}
Let $m$ be a reachable marking of an LPN $N$. 
We say that $m$ is \emph{honored} iff $m(s)\geq 0$
for all places $s$ of $N$.
\end{definition}

An honored firing sequence is a firing sequence where the final marking is honored. 
Note that if the net has no lending places, then all the reachable markings are honored.
 
\begin{example}
In the net of Ex.~\ref{ex:carl net}, the transition $\atom{c}$ is enabled 
even though there are no tokens in the places $p_2$ and $p_4$ in its pre-set, 
as they are lending places. 
The other transitions are not 
enabled, hence at the initial marking only $\atom{c}$ may be executed (on credit).
After firing $\atom{c}$, only $\atom{b}$ can be executed.
This results in putting one token in $p_3$ and one in $p_4$, hence giving back the one taken on credit. 
After this, only $\atom{a}$ can be executed.
Upon firing $\atom{c}$, $\atom{b}$ and $\atom{a}$, the marking is honored. 
The net is clearly a (correctly labeled) occurrence net. 
\end{example}

%\paragraph{Composing LPNs.}
We now introduce a notion of composition of LPNs. 
The idea is that the places with a label are
places in an \emph{interface} of the net (though we do not put any limitation on such places, as done instead
\emph{e.g.} in \cite{Aalst10cj}) and they never are initially marked. The labelled transitions of a net are
connected with the places bearing the same label of the other. 
 
\begin{definition} \label{de:composition}
Let 
$N = (\seq{S, T, F, \Gamma, \Lambda},$ $m_0, \mathcal{L})$ and
$N' = (\seq{S', T', F', \Gamma', \Lambda'},$ $m'_0,$ $\mathcal{L}')$
be two LPNs.
We say that $N, N'$ are \emph{compatible} whenever
$(a)$ they have the same set of labels,
$(b)$ $S\cap S' = \emptyset$, 
$(c)$ $T\cap T' = \emptyset$, 
$(d)$ $m_0(s) = 1$ implies $\Gamma(s) = \bot$, 
and
$(e)$ $m'_0(s') = 1$ implies $\Gamma'(s') = \bot$.
If $N$ and $N'$ are compatible, their composition
$N\oplus N'$ is the LPN
 $(\seq{\hat{S}, T \cup T', \hat{F}, \hat{\Gamma}, \hat{\Lambda}}, 
 \hat{m}_0,$ $\hat{\mathcal{L}})$ in Fig.~\ref{fig:lpn:oplus}.

\begin{figure}[t]
\[
\begin{array}{lcl}
  \hat{S} & = &
  \begin{array}{ll}
    (S \setminus \setcomp{s\in S}{\Gamma(s) \in \Lambda'(T') \text{ and } \post{s} = \emptyset}) \;\cup\; \\ 
    (S' \setminus \setcomp{s'\in S'}{\Gamma'(s') \in \Lambda(T) \text{ and } \post{s'} = \emptyset})
  \end{array}
  \\[15pt]
  \hat{F}(\hat{s},\hat{t}) & \iff &
  \big( \hat{s} = s_1 \in S \;\land\; \hat{t} = t_1\in T \;\land\; F(s_1,t_1) \big) \\
  & \;\lor\; & 
  \big( \hat{s} = s_2 \in S' \;\land\; \hat{t} = t_2\in T' \;\land\; F'(s_2,t_2) \big)
  \\[5pt]
  \hat{F}(\hat{t},\hat{s}) & \iff &
  \big( \hat{s} = s_1\in S \;\land\; \hat{t} = t_1\in T \;\land\; F(t_1,s_1) \big) \\
  & \;\lor\; &
  \big( \hat{s} = s_2 \in S' \;\land\; \hat{t} = t_2\in T' \;\land\; F'(t_2,s_2) \big)
  \\
  & \;\lor\; & 
  \big( \hat{s} = s\in S \;\land\; \hat{t} = t'\in T' \;\land\; \Lambda'(t') = \Gamma(s) \neq\bot \big)\\
  & \;\lor\; &
  \big( \hat{s} = s'\in S' \;\land\; \hat{t} = t\in T \;\land\; \Lambda(t) = \Gamma'(s') \neq\bot \big)
  \\[5pt]
  \hat{\Gamma}(\hat{s}) & = & \begin{cases}
                  \Gamma(s_1) & \mathit{if}\ \hat{s} = s_1\in S \\
                  \Gamma'(s_2) & \mathit{if}\ \hat{s} = s_2\in S'
                  \end{cases}
  \\[15pt]
  \hat{\Lambda}(\hat{t}) & = & \begin{cases}
                  \Lambda(t_1) & \mathit{if}\ \hat{t} = t_1\in T\\
                  \Lambda'(t_2) & \mathit{if}\ \hat{t} = t_2\in T'\\
                  \end{cases}
  \\[15pt]
  \hat{m}_0(\hat{s}) & = & \begin{cases}
    1 & \text{if $\hat{s} = s_1\in S$ and $m_0(s_1) = 1$, or $\hat{s} = s_2\in S'$ and $m'_0(s_2) = 1$} \\
    0 & \text{otherwise}
  \end{cases}
  \\[15pt]
  \hat{\mathcal{L}} & = & (\mathcal{L} \;\cup\; \mathcal{L}') \;\cap\; \hat{S}
\end{array}
\]
\vspace{-5pt}
\hrulefill
\vspace{-5pt}
\caption{Composition of two LPNs.}  \label{fig:lpn:oplus}
\end{figure}
% Then $N\oplus N'$ is the LPN $\corrlab{(\langle\hat{S}, \hat{T}, \hat{F}, \hat{\Gamma}, 
% \hat{\Lambda}\rangle, \hat{m}_0, \hat{\mathcal{L}})}$.
 \end{definition}

The underlying idea of LPN composition is rather simple: 
the sink places in a net bearing a label
of a transition of the other net are removed, and places and transitions with the same
label are connected accordingly 
(the removed sink places have places with the same label in the other net). 
All the other ingredients of the compound net  
are trivially inherited from the components.
Observe that, when composing two compatible nets $N$ and $N'$ such that 
$\Gamma(S)\cap\Gamma'(S') =\emptyset$, we obtain the disjoint union 
of the two nets. 
Further, if the common label $a\in \Gamma(S)\cap\Gamma'(S')$ is associated 
in $N$ to a place $s$ with  empty post-set and in $N'$ to a place $s'$ 
with empty post-set (or \emph{vice versa}) and the labelings
are injective, we obtain precisely the composition defined in~\cite{Aalst10cj}.  
If the components $N$ and $N'$ may satisfy some properties (sets of markings $\mathcal{M}$ and 
$\mathcal{M}'$), the compound net $N\oplus N'$ may satisfy the compound property (which is 
the set of markings $\hat{\mathcal{M}}$ obtained obviously from $\mathcal{M}$ and 
$\mathcal{M}'$).

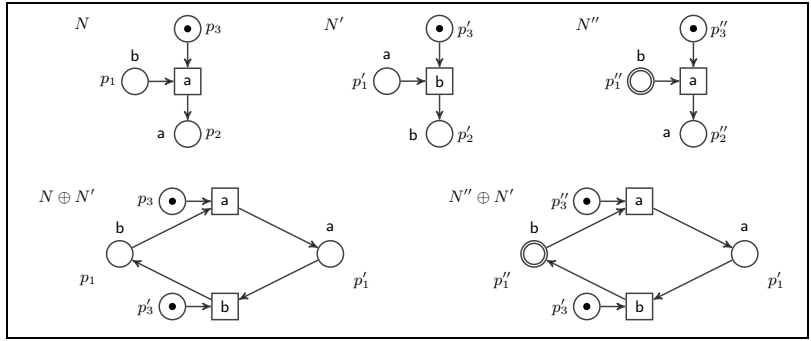
\begin{figure}[t]
\begin{center}
\fbox{
\begin{tabular}{c}
\scalebox{0.7}{ 
 \begin{tikzpicture}[>=stealth'] 
 \tikzstyle{lplace}=[circle,thick,draw=black!75,minimum size=4mm]
 \tikzstyle{invplace}=[circle,thick,draw=black!00,minimum size=5mm]
 \tikzstyle{place}=[circle,thick,draw=black!75,minimum size=5mm]
 \tikzstyle{transition}=[rectangle,thick,draw=black!75,minimum size=5mm]
 \tikzstyle{edge}=[->,thick,draw=black!75]
 \tikzstyle{edgered}=[->,thick,draw=red!75]
 \tikzstyle{edgeblu}=[->,thick,draw=blue!75]
 \path ( 0, 0) node [place] (p1) {}
       ( 0, 0) node [xshift=-5mm] {$p_1$}
       ( 0, 0) node [yshift=5mm] {\atom{b}}
       ( 1, 0) node [transition] (a) {\atom{a}}
       ( 1,-1) node [place] (p2) {}
       ( 1,-1) node [xshift=5mm] {$p_2$}
       ( 1,-1) node [xshift=-5mm] {\atom{a}}
       ( 1, 1) node [place] (p3) {}
       ( 1, 1) node [xshift=5mm] {$p_3$}
       ( 2, 0) node [invplace] {}
       ( 1, 1) node [token] {};
       \draw [edge] (a) to (p2);
       \draw [edge] (p1) to (a);
       \draw [edge] (p3) to (a);
       \node at (-1, 1.1) {$N$};
 \end{tikzpicture}
 \hspace{1cm}
 \begin{tikzpicture}[>=stealth'] 
 \tikzstyle{lplace}=[circle,thick,draw=black!75,minimum size=4mm]
 \tikzstyle{place}=[circle,thick,draw=black!75,minimum size=5mm]
 \tikzstyle{invplace}=[circle,thick,draw=black!00,minimum size=5mm]
 \tikzstyle{transition}=[rectangle,thick,draw=black!75,minimum size=5mm]
 \tikzstyle{edge}=[->,thick,draw=black!75]
 \tikzstyle{edgered}=[->,thick,draw=red!75]
 \tikzstyle{edgeblu}=[->,thick,draw=blue!75]
 \path ( 0, 0) node [place] (p1) {}
       ( 0, 0) node [xshift=-5mm] {$p'_1$}
       ( 0, 0) node [yshift=5mm] {\atom{a}}
       ( 1, 0) node [transition] (b) {\atom{b}}
       ( 1,-1) node [place] (p2) {}
       ( 1,-1) node [xshift=5mm] {$p'_2$}
       ( 1,-1) node [xshift=-5mm] {\atom{b}}
       ( 1, 1) node [place] (p3) {}
       ( 2, 0) node [invplace] {}
       ( 1, 1) node [xshift=5mm] {$p'_3$}
       ( 1, 1) node [token] {};
       \draw [edge] (b) to (p2);
       \draw [edge] (p1) to (a);
       \draw [edge] (p3) to (a);
       \node at (-1, 1.1) {$N'$};
 \end{tikzpicture}
 \hspace{1cm}
 \begin{tikzpicture}[>=stealth'] 
 \tikzstyle{lplace}=[circle,thick,draw=black!75,minimum size=4mm]
 \tikzstyle{place}=[circle,thick,draw=black!75,minimum size=5mm]
 \tikzstyle{invplace}=[circle,thick,draw=black!00,minimum size=5mm]
 \tikzstyle{transition}=[rectangle,thick,draw=black!75,minimum size=5mm]
 \tikzstyle{edge}=[->,thick,draw=black!75]
 \tikzstyle{edgered}=[->,thick,draw=red!75]
 \tikzstyle{edgeblu}=[->,thick,draw=blue!75]
 \path ( 0, 0) node [place] (p1) {}
       ( 0, 0) node [xshift=-5mm] {$p''_1$}
       ( 0, 0) node [yshift=5mm] {\atom{b}}
       ( 0, 0) node [lplace] {}
       ( 1, 0) node [transition] (a) {\atom{a}}
       ( 1,-1) node [place] (p2) {}
       ( 1,-1) node [xshift=5mm] {$p''_2$}
       ( 1,-1) node [xshift=-5mm] {\atom{a}}
       ( 1, 1) node [place] (p3) {}
       ( 1, 1) node [xshift=5mm] {$p''_3$}
       ( 1, 1) node [token] {}
       ( 2, 0) node [invplace] {};
       \draw [edge] (a) to (p2);
       \draw [edge] (p1) to (a);
       \draw [edge] (p3) to (a);
       \node at (-1, 1.1) {$N''$};
 \end{tikzpicture}} 
\\[10pt]
\scalebox{0.7}{
 \begin{tikzpicture}[scale=1,>=stealth'] 
 \tikzstyle{lplace}=[circle,thick,draw=black!75,minimum size=4mm]
 \tikzstyle{place}=[circle,thick,draw=black!75,minimum size=5mm]
 \tikzstyle{transition}=[rectangle,thick,draw=black!75,minimum size=5mm]
 \tikzstyle{edge}=[->,thick,draw=black!75]
 \tikzstyle{edgered}=[->,thick,draw=red!75]
 \tikzstyle{edgeblu}=[->,thick,draw=blue!75]
 \path ( 0, 1) node [place] (p1) {}
       ( 0, 1) node [xshift=-6mm,yshift=-5mm] {$p_1$}
       ( 0, 1) node [yshift=5mm] {\atom{b}}
       ( 2, 2) node [transition] (a) {\atom{a}}
       ( 4, 1) node [place] (p2) {}
       ( 4, 1) node [xshift=6mm,yshift=-5mm] {$p'_1$}
       ( 4, 1) node [yshift=5mm] {\atom{a}}
       ( 1, 2) node [place] (p3) {}
       ( 1, 2) node [xshift=-5mm] {$p_3$}
       ( 2, 0) node [transition] (b) {\atom{b}}
       ( 1, 2) node [token] {}
       ( 1, 0) node [place] (p4) {}
       ( 1, 0) node [xshift=-5mm] {$p'_3$}
       ( 1, 0) node [token] {};       
       \draw [edge] (a) to (p2);
       \draw [edge] (p1) to (a);
       \draw [edge] (p3) to (a);
       \draw [edge] (p2) to (b);
       \draw [edge] (b) to (p1);
       \draw [edge] (p4) to (b);
       \node at (-1,2.1) {$N\oplus N'$};
 \end{tikzpicture}
\hspace{30pt}
 \begin{tikzpicture}[>=stealth'] 
 \tikzstyle{lplace}=[circle,thick,draw=black!75,minimum size=4mm]
 \tikzstyle{place}=[circle,thick,draw=black!75,minimum size=5mm]
 \tikzstyle{transition}=[rectangle,thick,draw=black!75,minimum size=5mm]
 \tikzstyle{edge}=[->,thick,draw=black!75]
 \tikzstyle{edgered}=[->,thick,draw=red!75]
 \tikzstyle{edgeblu}=[->,thick,draw=blue!75]
 \path ( 0, 1) node [place] (p1) {}
       ( 0, 1) node [xshift=-6mm,yshift=-5mm] {$p''_1$}
       ( 0, 1) node [yshift=5mm] {\atom{b}}
       ( 0, 1) node [lplace] {}
       ( 2, 2) node [transition] (a) {\atom{a}}
       ( 4, 1) node [place] (p2) {}
       ( 4, 1) node [xshift=6mm,yshift=-5mm] {$p'_1$}
       ( 4, 1) node [yshift=5mm] {\atom{a}}
       ( 1, 2) node [place] (p3) {}
       ( 1, 2) node [xshift=-5mm] {$p''_3$}
       ( 1, 2) node [token] {}
       ( 2, 0) node [transition] (b) {\atom{b}}
       ( 1, 0) node [place] (p4) {}
       ( 1, 0) node [xshift=-5mm] {$p'_3$}
       ( 1, 0) node [token] {};       
       \draw [edge] (a) to (p2);
       \draw [edge] (p1) to (a);
       \draw [edge] (p3) to (a);
       \draw [edge] (p2) to (b);
       \draw [edge] (b) to (p1);
       \draw [edge] (p4) to (b);
       \node at (-1,2.1) {$N'' \oplus N'$};
 \end{tikzpicture}
}
\end{tabular}}
\end{center}
\vspace{-15pt}
\caption{Three LPNs (top) and their pairwise compositions (bottom).} 
\label{fig:a e b che si implicano a vicenda}
\end{figure}

\begin{example}\label{ex:a e b che si implicano a vicenda}
Consider the nets in Fig.~\ref{fig:a e b che si implicano a vicenda}.
Net $N$ fires $\atom{a}$ after $\atom{b}$ has been performed;
dually, net $N'$ waits for $\atom{b}$ before firing $\atom{a}$.
These nets model two participants which protect themselves by waiting the other one
to make the first step (the properties being that places $p_3$ and $p'_3$, respectively, 
are not marked). 
Clearly, no agreement is possible in this scenario.
This is modelled by the deadlock in the composition $N \oplus N'$, 
where neither transitions $\atom{a}$ nor $\atom{b}$ can be fired.
Consider now the LPN $N''$, which differs from $N$ only for the lending place $p_1''$.
This models a participant which may fire $\atom{a}$ on credit, 
under the \emph{guarantee} that the credit will be eventually honoured 
by the other participant performing $\atom{b}$
(hence, the participant modeled by $N''$ is still protected), and the property is then 
place $p_3''$ unmarked and $p_1''$ with a non negative marking.
The composition $N'' \oplus N'$ weakly terminates wrt the above properties, because
transition $\atom{a}$ can take a token on credit from $p''_1$, 
and then transition $\atom{b}$ can be fired, 
so honouring the debit in $p''_1$.
\end{example}

The operation $\oplus$ is clearly associative and commutative.
 
 \begin{myproposition}\label{prop:oplus commutativo ed associativo}
 Let $N_1$, $N_2$ and $N_3$ be three compatible LPNs. 
 Then, $N_1\oplus N_2 = N_2\oplus N_1$ and 
 $N_1\oplus (N_2\oplus N_3) = (N_1\oplus N_2)\oplus N_3$.
 \end{myproposition} 
 
The composition $\oplus$ does not have the property that, in general, 
considering only the transitions of
one of the components, we obtain the LPN we started with, 
\emph{i.e.} $(N_1\oplus N_2)|_{T_i} \neq N_i$.
This is because the number of places with labels increases and 
new arcs may be added, and these places are not
\emph{forgotten} when considering the subnet generated by $T_i$. 
However these added places are not initially marked, 
hence it may be that the nets have the same traces. 
 
\begin{definition}\label{de:net-equiv}
Let $N$ and $N'$ two LPNs on the same sets of labels. 
We say that $N$ \emph{approximates} $N'$ ($N \lesssim N'$) 
iff $\mathit{Traces}(N) \subseteq \mathit{Traces}(N')$. 
% We say $N,N'$ \emph{equivalent} 
We write $N \sim N'$ when $N \lesssim N'$ and $N' \lesssim N$.

\end{definition}
% Hence it would be nice to have that $(N_1\oplus N_2)|_{T_i} \sim N_i$.  Unfortunately this is not in 
% general true, though $(N_1\oplus N_2)|_{T_i}$ and 
% $N_i$ are \emph{almost} the same net: all the traces of $N_i$ are also traces of $(N_1\oplus N_2)|_{T_i}$ 
% (which may have some more traces as in the composition some place may become 
% \emph{lending}).
 
 \begin{myproposition}
 For two compatible LPNs $N_1, N_2$, 
 $N_i \sim (N_1\oplus N_2)|_{T_i}$, $i = 1, 2$. 
 \end{myproposition}

Following~\cite{Aalst10cj} we introduce a notion of refinement
(called \emph{accordance} in~\cite{Aalst10cj})
between two LPNs. 
% An example is in Fig.~\ref{fig:refinement}.
We say that $M$ (with a property $\mathcal{M}_{M}$) 
is a \emph{strategy} for an LPN $N$ (with a property $\mathcal{M}$)
if $N\oplus M$ is weakly terminating. 
With $\mathcal{S}(N)$ we denote the set of all strategies for $N$. 
In the rest of the paper we assume that properties are always specified, 
even when not done explicitly.

\begin{definition}\label{de:accordance}
An LPN $N'$ \emph{refines} $N$ if $\mathcal{S}(N') \supseteq \mathcal{S}(N)$.
\end{definition}
%
% As noticed in~\cite{Aalst10cj}, accordance defines a refinement relation 
% between nets, in this case lending Petri nets. 

% \begin{proposition}
Observe that if $N'$ refines $N$ and $N$ weakly terminates, 
% in $\mathcal{M}$ (a property of the executions of $N$),
then $N'$ weakly terminates as well. 

If a weakly terminating LPN $N$ is obtained by composition of several nets, 
\emph{i.e.} $N = \bigoplus_{i} N_i$, 
% where $I$ is a set of indexes, 
we can ask what happens if there is an $N'_i$ which refines $N_i$,
for each $i$. 
The following theorem gives the desired answer.
 
\begin{mythm}\label{th:weakly term and accordance}
Let $N = \bigoplus_{i} N_i$ be a weakly terminating LPN, 
and assume that $N'_i$ refines $N_i$, for all $i$. 
Then, $N' = \bigoplus_{i} N'_i$ is a weakly terminating LPN.
 \end{mythm}

The theorem above gives a compositional criterion to check weak termination
of a SOC application. 
One starts from an abstract specification (e.g.\ a choreography), 
projects it into a set of local views, and then refines each of them into
a service implementation.
These services can be verified independently (for refinement), 
and it is guaranteed that their composition still enjoys weak termination.

We now define, starting from a marking $m$,
which actions may be performed immediately after,
while preserving the ability to reach an honored marking.
We call these actions \emph{urgent}.

\begin{definition}\label{de:urg-mark}
For an LPN $N$ and marking $m$, %$m\in\mathsf{M}(N)$,
we say $\atom{a}$ \emph{urgent at} $m$ iff 
there exists a firing sequence $m\trans{t_1} \cdots \trans{t_n}m_n$
with $\Lambda(t_1) = \atom{a}$ and $m_n$ honored.
\end{definition}

\begin{example}
Consider the nets in Ex.~\ref{ex:a e b che si implicano a vicenda}.
In $N''\oplus N'$ 
the only urgent action at the initial marking is $\atom{a}$, while
$\atom{b}$ is urgent at the marking where $p_1'$ is marked. 
In $N''$ there are no urgent actions at the initial marking,
since no honored marking is reachable.
In the other nets ($N$, $N'$, $N \oplus N'$) 
no actions are urgent in the initial marking, since these nets
are deadlocked.
\end{example}

\section{Physical contracts}\label{sec:lnp-contracts}

We now present a model for physical contracts based on LPNs.
Let $\atom{a}, \atom{b}, \ldots \in \atoms$ be \emph{actions}, 
performed by \emph{participants} ${\pmv A}, {\pmv B}, \ldots \in \Princ$.
We assume that actions may only be performed once.
Hence, we consider a subclass of LPNs, namely occurrence nets, 
where all the transitions with the same label
are mutually exclusive. 
A physical contract is an LPN, together with 
a set $\aname$ of participants bound by the contract,
a mapping $\princsym$ 
from actions to participants,
and a set $\Omega$ modeling the states
where all the participant in $\aname$ are satisfied.

\begin{definition}\label{de:lpn-contracts}
A \emph{contract net} $\pname$ is a tuple 
$(O, \aname, \princsym, \Omega)$, where
$O$ is an occurrence LPN 
$(\langle S, T, F,$ $\Gamma, \Lambda\rangle,$ $m_0, \mathcal{L})$
labeled on $\mathcal{T}$,
$\aname \subseteq \Princ$,
$\princsym: \mathcal{T} \rightarrow \Princ$,
$\payoffsym \subseteq \powset{\atoms}$ is the set of goals of the participants,
and where:
\begin{enumerate}[(a)]

\item
$\forall s \in S.\; 
(m_0(s) = 1 \implies \pre{s} = \emptyset \;\land\; \Gamma(s) = \bot) \;\land\;
(s\in\mathcal{L} \implies \Gamma(s)\in\mathcal{T})$,
 %\item $\forall t\in T$. $\exists s\in \pre{t}$ and $\Gamma(s)$ is undefined,

\item 
$\forall t\in T.\;
(\forall s\in\post{t}. \; \Lambda(t) = \Gamma(s)) \;\land\;
(\exists s\in\pre{t}.  \; s \not\in \mathcal{L})$,
% each transition has at least a non-lending place in its preset

\item 
$\forall t, t'\in T$. $\Lambda(t) = \Lambda(t') \implies \exists s\in\pre{t}\cap\pre{t'}.\; m_0(s) = 1$,

% \item 
%  $\forall s\in\mathcal{L}$. $\Gamma(s)\in\mathcal{T}$,
% % NdM: this implies that lending places are always labeled

\item \label{de:lpn-contracts:princ-aname} 
$\princ{\Lambda(T)} \subseteq \aname$.

\end{enumerate}
% and where $\aname \subseteq \Princ$,
% $\princsym: \mathcal{T} \rightarrow \aname$,
% and $\Omega$ is a set of subsets of $\atoms$.
\end{definition}

The last constraint
% ~\eqref{de:lpn-contracts:princ-aname} 
models the fact that only the participants in
$\aname$ may perform actions in $\pname$.

Given a state $X$ of the component $O$ of $\pname$,
% $O = (\langle S, T, F,$ $\Gamma, \Lambda\rangle, m_0, \mathcal{L})$ 
% which is a part of a contract net $\pname$, 
the reached marking $m$ tells us which actions have been performed,
% (not which transitions have been executed) 
and which tokens have been taken on credit.
The \emph{configuration} $\mu(m)$
associated to a marking $m$ is the pair $(C,Y)$ defined as:
\begin{itemize}
  \item $C = \setcomp{\atom{a}\in \mathcal{T}}{\exists s\in S.\; \setenum{s} = 
         \bigcap_{t\in T}  \setcomp{\pre{t}}{\Lambda(t) = \atom{a}} 
         \;\mathit{and}\; m(s) = 0}$, and  
  \item $Y = \setcomp{\atom{a}\in \mathcal{T}}{\exists s\in S. \; 
         \atom{a} = \Gamma(s)\ \mathit{and}\ m(s) < 0}$
\end{itemize}
The first component is the set of the labels of the transitions in $X$.
The marking $m$ is honored whenever the second component of $\mu(m)$ is empty.

We now state the conditions under which two contract nets can be composed. 
We require that an action can be performed only by one of the components 
(the other may \emph{use} the tokens produced by the execution of such action).

\begin{definition}\label{de:contract net compatible}
Two contracts nets $\pname = (O, \aname, \princsym, \Omega)$ and
$\pname' = (O', \aname', \princsym',$ $\Omega')$ are \emph{compatible} 
whenever $O\oplus O'$ is defined and 
$\aname \cap \aname' = \emptyset$.
% $\Lambda(T)\cap\Lambda'(T') = \emptyset$. 
\end{definition}

\noindent
The composition of $\pname$ and $\pname'$ is then the obvious extension of the one on LPNs:
\begin{definition}\label{de:comp of contract nets}
Let $\pname = (O, \aname, \princsym, \Omega)$ and
$\pname' = (O', \aname', \princsym',$ $\Omega')$ be two compatible contract nets. Then
$\pname\oplus\pname' = (O\oplus O', \aname\cup\aname', \princsym\circ\princsym', \Omega'')$ where  
$\Omega'' = \setenum{X\cup X'\ |\ X\in\Omega, X'\in\Omega'}$.
\end{definition}

% We use the same symbol to denote the composition of LPN nets and of contract nets, being the latter based on the former.

We lift the notion of weak termination to contract nets 
$\pname = (O, \aname, \princsym, \oksym, \Omega)$.
The set of markings obtained by $\Omega$ is 
$\mathcal{M}_{\Omega} = \setcomp{m\in \mathsf{M}(O)\!\!}{\!\mu(m) = (C,\emptyset),\ C\in\Omega}$.
We say that $\pname$ weakly terminates w.r.t.\ $\Omega$ when $O$ weakly 
terminates w.r.t.\ $\mathcal{M}_{\Omega}$.

We also extend to contract nets the notion of urgent actions given for LPNs 
(Def.~\ref{de:urg-mark}).
Here, the set of urgent actions $\curgent[\pname]{C}$ is parameterized 
by the set $C$ of actions already performed.

\begin{definition}\label{de:urg-contracts}
Let 
$\pname$ 
be a contract net, and let $C\subseteq \mathcal{T}$. 
We define:
\[
  \curgent[\pname]{C} 
   = 
  \setcomp{\atom{a}\in\mathcal{T}}
  {\exists Y \subseteq \atoms.\; 
   \exists m.\ %\in\mathsf{M}(O) 
   \mu(m) = (C,Y)\ \;\land\; \atom{a}\ \text{ is urgent at }\ m}
\]
\end{definition}

\begin{example}
Interpret the LPN $N_1'$ in Fig.~\ref{fig:carl net} as a contract net where 
the actions $\atom{a}$, $\atom{b}$, $\atom{c}$ are associated, respectively, 
to participants $\pmv{A}$, $\pmv{B}$, and $\pmv{C}$, and $\Omega$ is immaterial. 
Then, $\atom{a}$ and $\atom{c}$ are urgent at the initial marking, whereas
$\atom{b}$ is not (the token borrowed from $p_1$ cannot be given back).
In the state where {\atom a} has been fired, only {\atom b} is urgent;
in the state where {\atom c} has been fired, no actions are urgent.
\end{example}

\section{Logical contracts} \label{sec:pcl}

In this section we briefly review Propositional Contract Logic 
(\pcl~\cite{BZ10lics}),
and we exploit it to model contracts.
\pcl extends intuitionistic propositional logic IPC
with a connective~$\coimp$, called \emph{contractual implication}.
Intuitively,
% differently from IPC,
a formula $\atom{b} \coimp \atom{a}$ implies $\atom{a}$ 
not only when $\atom{b}$ is true, 
like IPC implication, but also in the case that a ``compatible'' 
formula, e.g.\ $\atom{a} \coimp \atom{b}$, holds.
\pcl allows for a sort of ``circular'' assume-guarantee reasoning,
hinted by 
$(\atom{b} \coimp \atom{a}) \land (\atom{a} \coimp \atom{b}) \imp \atom{a} \land \atom{b}$,
which is a theorem in \pcl.
We assume that the prime formulae of \pcl coincide with the atoms in $\atoms$.
% The symbol $\top$ denotes ``true''.
\pcl formulae, ranged over greek letters $\varphi, \varphi',\ldots$,
% \begin{definition} \label{def:pcl:syntax}
% The formulae of \pcl\ 
are defined as:
\[
  \varphi \; ::= \;
  \bot \;\mid\; \top \;\mid\; \atom{a} \;\mid\; \lnot \varphi \;\mid\; 
  \varphi \lor \varphi \;\mid\; \varphi\land \varphi \;\mid\; \varphi \imp \varphi \;\mid\; \varphi \coimp \varphi
\]
% We let $p\leftrightarrow q$ be syntactic sugar for $(p\imp q)\land(q\imp p)$.
% If a formula is $\coimp$-free, we say it is an IPC formula.
% \end{definition}

\begin{figure}[t]
\[
% \normalsize
\small
  \irule
  {\begin{array}{c} \\ \Delta \vdash \psi \end{array}}
  {\Delta \vdash \varphi \coimp \psi}
  \begin{array}{c} \\ \;\nrule{($\coimp$I1)} \end{array}
  \qquad
  \irule
  {\begin{array}{c} \\ \Delta \vdash \varphi' \coimp \psi' \end{array} \quad
   \begin{array}{rl}
     \Delta,\varphi \, & \vdash \varphi' \\[2pt]
     \Delta,\psi' & \vdash \varphi \coimp \psi
   \end{array}}
 {\Delta \vdash \varphi \coimp \psi}
 \begin{array}{c} \\ \;\nrule{($\coimp$I2)} \end{array}
 \qquad
  \irule
  {\begin{array}{rl} 
      \Delta & \vdash \varphi \coimp \psi \\[2pt]
      \Delta, \psi & \vdash \varphi
    \end{array}}
  {\Delta \vdash \psi}
  \begin{array}{c} \\ \;\nrule{($\coimp$E)} \end{array}
\]
\hrulefill
\vspace{-10pt}
\caption{Natural deduction for \pcl\ (rules for $\coimp$).}
\label{fig:pcl:nd}
\end{figure}

Two proof systems have been presented for \pcl\!:
a sequent calculus~\cite{BZ10lics}, and an equivalent
natural deduction system~\cite{BCDZ13ice},
the main rules of which are shown in Fig.~\ref{fig:pcl:nd}.
Provable formulae are contractually implied,
according to rule~\nrule{($\coimp$I1)}.
Rule~\nrule{($\coimp$I2)} provides $\coimp$ with the same weakening properties
of $\imp$.
The crucial rule is \nrule{($\coimp$E)}, 
which allows for the elimination of $\coimp$.
Compared to the rule for elimination of $\imp$ in IPC, 
the only difference is that
in the context used to deduce the antecedent $\varphi$, 
rule \nrule{($\coimp$E)} also allows for using 
as hypothesis the consequence~$\psi$.
The decidability of the provability relation of \pcl
has been proved in~\cite{BZ10lics}, by exploiting the cut elimination
property enjoyed by the sequent calculus.

\smallskip
To model contracts, we consider the Horn fragment of \pcl\!, 
which comprises atoms, conjunctions, and non-nested 
(intuitionistic/contractual) implications.

\begin{definition} \label{def:pcl-contract}
% Let Horn \pcl formulae $\varphi, \varphi', \ldots$ be defined as follows:
% \[
% \textstyle
%   \varphi \; ::= \; \bigwedge_{i \in \mathcal{I}} \; \alpha_i 
%   \hspace{40pt}
%   \alpha \; ::= \; \bigwedge_{j \in \mathcal{J}} \; \atom{b}_j 
%         \ \ \big| \ \ 
%     ( \bigwedge_{i \in \mathcal{I}} \; \atom{a}_i ) \imp \atom{b}
%         \ \ \big| \ \ 
%    ( \bigwedge_{i \in \mathcal{I}} \; \atom{a}_i ) \coimp \atom{b}
% \]
A \pcl contract is a tuple $\seq{\Delta,\aname,\princsym,\payoffsym}$,
where $\Delta$ is a Horn \pcl\ theory,
$\aname \subseteq \Princ$,
$\princsym: \atoms \rightarrow \Princ$ associates each atom with a participant,
and
$\payoffsym \subseteq \powset{\atoms}$ is the set of goals
of the participants.
\end{definition}

The component $\aname$ of $\cname$ 
contains the participants which can promise to do something in $\cname$.
Consequently, we shall only consider \pcl\ contracts such that
if $\alpha \circ \atom{a} \in \Delta$, 
for $\circ \in \setenum{\imp,\coimp}$,
then $\princ{\atom{a}} \in \aname$.
% Indeed, a clause $p \circ \atom{a}$ models the promise to do action~{\atom a}
% (under the given premises).
% provided that the premises $p$ have been performed,
% The requirement above asks that if a contract promises to do $\atom{a}$
% (under whatever premises), then $\atom{a}$ must be an action of some
% participant in $\aname$.

\begin{example} \label{ex:toys:contract}
Suppose three kids want to play together.
Alice has a toy airplane, Bob has a bike, and Carl has a toy car.
Each of the kids is willing to share his toy, but they have different constraints:
Alice will lend her airplane only \emph{after} Bob has 
allowed her ride his bike;
Bob will lend his bike after he has played with Carl's car;
Carl will lend his car if the other two kids promise to
eventually let him play with their toys.
Let % $\aname = \setenum{\pmv A, \pmv B, \pmv C}$, and let
$\princsym = \setenum{\atom{a} \mapsto {\pmv A}, \atom{b} \mapsto {\pmv B}, \atom{c} \mapsto {\pmv C}}$.
The kids contracts are modeled as follows:
%\[
  % \cname_{\pmv A} = 
  $\seq{\atom{b} \imp \atom{a}, \setenum{\pmv A}, \princsym, \setenum{\setenum{\atom{b}}}}$, 
%  \qquad
  % \cname_{\pmv B} = 
  $\seq{\atom{c} \imp \atom{b}, \setenum{\pmv B}, \princsym, \setenum{\setenum{\atom{c}}}}$, and 
%  \qquad
  % \cname_{\pmv C} = 
  $\seq{(\atom{a} \land \atom{b}) \coimp \atom{c}, \setenum{\pmv C}, \princsym, \setenum{\setenum{\atom{a},\atom{b}}}}$.
%\]
\end{example}

A contract admits an \emph{agreement} when all the involved 
participants can reach their goals.
This is formalized in Def.~\ref{def:agreement} below.

\begin{definition} \label{def:agreement}
A \pcl contract admits an \emph{agreement} iff 
$\exists X \in \payoffsym.\; \Delta \vdash \bigwedge X$.
\end{definition}

We now define composition of \pcl contracts.
If $\cname'$ is the contract of an adversary of $\cname$, 
then a na\"ive composition of the two contracts could easily lead to an attack,
e.g.\ when Mallory's contract says that Alice is obliged to give him
her airplane.
To prevent from such kinds of attacks,
contract composition is a partial operation.
We do \emph{not} compose contracts which bind the same participant,
or which disagree on the association between atoms and participants.

\begin{definition}
Two \pcl contracts 
$\cname = \seq{\Delta,\aname,\princsym,\payoffsym}$ and
$\cname' = \seq{\Delta',\aname',\princsym',\payoffsym'}$
are \emph{compatible} whenever $\aname \cap \aname' = \emptyset$, and
\(
  \forall {\pmv A} \in \aname \cup \aname'.\;\;
  \invprinc{\pmv A} = \princsym'^{\, -1}({\pmv A})
\).
If $\cname$, $\cname'$ are compatible, the contract 
$\cname \ccomp \cname' = \seq{\Delta \cup \Delta', \aname \cup \aname', \princsym \circ \princsym', \payoffsym \mid \payoffsym'}$, where
$\payoffsym \mid \payoffsym' = \setcomp{X \cup X'}{X \in \payoffsym,\; X' \in \payoffsym'}$, is their composition.
\end{definition}

\begin{example} \label{ex:toys:contract-composition}
The three contracts in Ex.~\ref{ex:toys:contract} are compatible, and
their composition is
\(
  \cname
  =
  \seq{\Delta, \setenum{{\pmv A}, {\pmv B}, {\pmv C}}, \setenum{\atom{a} \mapsto {\pmv A}, \atom{b} \mapsto {\pmv B}, \atom{c} \mapsto {\pmv C}}, \setenum{\setenum{\atom{a},\atom{b},\atom{c}}}}
\)
where $\Delta$ is the theory
\(
  \setenum{ 
  \atom{b} \imp \atom{a},\;
  \atom{c} \imp \atom{b},\;
  (\atom{a} \land \atom{b}) \coimp \atom{c}
  }
\).
$\cname$ has an agreement, since
$\Delta \vdash \atom{a} \land \atom{b} \land \atom{c}$.
The agreement exploits the fact that Carl's contract allows 
the action $\atom{c}$ to happen ``on credit'', before the other actions
are performed.
\end{example}

We now recap from~\cite{BCDZ13ice} the notion of \emph{proof traces},
i.e.\ the sequences of atoms respecting the order imposed by proofs in \pcl\!.
Consider e.g.\ rule \nrule{($\imp$E)}:
\[
  \irule
  {\Delta \vdash \alpha \imp {\atom a} \qquad \Delta \vdash \alpha}
  {\Delta \vdash {\atom a}}
  \;\nrule{($\imp$E)}
\]
The rule requires a proof of all the atoms in $\alpha$ in order to 
construct a proof of~{\atom a}.
% Assuming to associate each ordering with a trace of atoms,
Accordingly, if $\sigma$ is a proof trace of $\Delta$,
then $\sigma a$ if a proof trace of $\Delta$.
Instead, in the rule \nrule{($\coimp$E)},
the antecedent $\alpha$ needs not necessarily be proved before {\atom a}:
it suffices to prove $\alpha$ by taking $a$ as hypothesis.

\begin{definition}[Proof traces~\cite{BCDZ13ice}]
\label{def:pcl:proof-trace}
For a Horn \pcl\ theory $\Delta$, 
we define the set of \emph{proof traces} $\sem{\Delta}$
by the rules in Fig.~\ref{fig:pcl:proof-traces},
where for $\sigma,\eta \in E^*$ we denote with
$\mkset{\sigma}$ the set of atoms in $\sigma$, with
$\sigma \eta$ the concatenation of $\sigma$ and $\eta$,
and with $\sigma \mid \eta$ the interleavings 
of $\sigma$ and $\eta$.
We assume that both concatenation and interleaving remove duplicates from the right,
e.g.\ $aba \mid ca = ab \mid ca = \setenum{abc,acb,cab}$.
\end{definition}

\begin{figure}[t]
\[
\begin{array}{c}
  \irule
  {}
  {\epsilon \in \sem{\Delta}}
  \,\nrule{($\epsilon$)}
  \hspace{12pt}
  \irule
  {\alpha \imp {\atom a} \in \Delta \quad
   \sigma \in \sem{\Delta} \quad 
   \mkset{\alpha} \subseteq \mkset{\sigma}}
  {\sigma \, {\atom a} \in \sem{\Delta}}
  \,\nrule{($\imp$)}
  \hspace{12pt}
  \irule
  {\alpha \coimp {\atom a} \in \Delta \quad 
   \sigma \in \sem{\Delta, {\atom a}} \quad
   \mkset{\alpha} \subseteq \mkset{\sigma}}
  {\sigma \mid {\atom a} \,\subseteq\, \sem{\Delta}}
  \,\nrule{($\coimp$)}
\end{array}
\]
\hrulefill
\vspace{-10pt}
\caption{Proof traces of Horn \pcl\!.}
\label{fig:pcl:proof-traces}
\end{figure}

The set $\curgent[\cname]{X}$ in Def.~\ref{def:pcl:urgent}
contains, given a set $X$ of atoms, 
the atoms which may be proved immediately after,
following some proof trace of $\cname$. 

\begin{definition}[Urgent actions~\cite{BCDZ13ice}] \label{def:pcl:urgent}
For a contract $\cname = \seq{\Delta,\ldots}$ and 
a set of atoms $X$, we define
% the set of atoms 
\(
  \curgent[\cname]{X}
  \; = \;
  \setcomp{{\atom a} \not\in X}
  {
  \exists \sigma, \sigma'. \;\;
  \mkset{\sigma} = X  
  \;\land\;
  \sigma \, {\atom a} \, \sigma' \in \sem{\Delta, X}
  }
\).
\end{definition}

\begin{example} \label{ex:pcl:urgent}
For the contract $\cname$ specified by the theory
$\Delta = {\atom a} \imp {\atom b},\, {\atom b} \coimp {\atom a}$,
we have 
$\sem{\Delta} = \setenum{\epsilon, {\atom a}{\atom b}}$,
and
$\curgent[\Delta]{\emptyset} = \setenum{\atom a}$,
$\curgent[\Delta]{\setenum{\atom a}} = \setenum{\atom b}$, 
$\curgent[\Delta]{\setenum{\atom b}} = \setenum{\atom a}$, and 
$\curgent[\Delta]{\setenum{{\atom a},{\atom b}}} = \emptyset$.
\end{example}

\section{From logical to physical contracts} \label{sec:translation}

In this section we show, starting from a logical contract, 
how to construct a physical one which preserves the agreement property.
Technically, we shall relate provability in \pcl to reachability of 
suitable configurations in the associated LPN.
The idea of our construction is to translate each Horn clause of a \pcl formula into 
a transition of an LPN, labelled with the action in the conclusion of the clause.

\begin{figure}[t]
\[
\begin{array}{lcl}
  S & = & (\mathcal{T}\times T) \cup \big( (\setcomp{\atom{a}}{\bigwedge X \imp \atom{a}\in\Delta} 
           \cup \setcomp{\atom{a}}{\bigwedge X \coimp \atom{a}\in\Delta} \cup \setcomp{\atom{a}}{\atom{a}\in\Delta}) \times \setenum{\ast} \big) 
  \\[5pt]
  T & = & \setcomp{(X,\atom{a},\nolending)}{\bigwedge X \imp \atom{a}\in\Delta} \;\cup\; 
  \setcomp{(X,\atom{a},\lending)}{\bigwedge X \coimp \atom{a}\in\Delta}
  \\[5pt]
  F & = & 
  \setcomp{(s,t)}{s = (\atom{a},\ast),\ t = (X,\atom{a},z)} 
  \;\cup\; 
  \setcomp{(s,t)}{s = (\atom{a},t),\ t = (X,\atom{c},z),\ \atom{a} \in X}
  \;\cup\;
  \\
  & & 
  \setcomp{(t,s)}{s = (\atom{a},x),\ t = (X,\atom{a},z),\ x \neq \ast}
  \\[5pt]
  \Gamma(s) & = & \wif{s = (\atom{a},x) \text{ with } x\in T}{\atom{a}}{\bot} 
  \\[5pt]
  % \Lambda : T \to \mathcal{T}$ such that 
  \Lambda(t) & = & \wif{t = (X,\atom{a},z)}{\atom{a}}{\bot}
  \\[5pt]
  m_0(s) & = & \wif{s = (\atom{a},\ast)}{1}{0}
  \\[5pt]
  \mathcal{L} & = & \setcomp{s\in S}{s = (\atom{a},t)\ \mathit{and}\ t = (X,\atom{c},\lending)\ 
        \mathit{with}\ X\neq\emptyset}
  \\
\end{array}
\]
\hrulefill
\vspace{-10pt}
\caption{Translation from logical to physical contracts.} 
\label{fig:pnl-contracts}
\end{figure}

\begin{definition}\label{de:pnl-contracts}
Let $\cname = \seq{\Delta,\aname,\princsym,\payoffsym}$ be a \pcl contract.
We define the contract net $\contrtopnl(\cname)$ as 
$((\langle S, T, F, \Gamma, \Lambda\rangle, m_0, \mathcal{L}), \aname, \princsym, \Omega)$
in Fig.~\ref{fig:pnl-contracts}.
\end{definition}

The transitions associated to $\cname$ are a subset $T$ of 
$\powset{\mathcal{T}}\times\mathcal{T}\times\setenum{\lending,\nolending}$.
For each intuitionistic/contractual implication, we introduce a transition as follows.
A clause $\bigwedge X \coimp \atom{a}$ maps to $(X,\atom{a},\lending) \in T$, 
while $\bigwedge X \imp \atom{a}$ maps to $(X,\atom{a},\nolending) \in T$. 
A formula $\atom{a}$ is dealt with as the clause
$\bigwedge \emptyset \imp \atom{a}$.
Places in $S$ carry the information on which transition may actually put/consume a token from them (even on credit). 
The lending places are those places $(\atom{a},t)$ where $t = (X,\atom{c},\lending)$. 
Observe that a transition $t = (X,\atom{a},z)$ puts a token in each place $(\atom{a},x)$ with $x\neq \ast$, 
and all the transitions bearing the same labels, say $\atom{a}$, 
are mutually excluding each other, as they share the unique input place $(\atom{a},\ast)$.
The initial marking will contains all the places in $\atoms \times\setenum{\ast}$, and if a token is consumed from one of these places then the place will be never marked again. 
Furthermore the lending places are never initially marked. 

\begin{figure}[b]
\begin{center}
\fbox{
\begin{tabular}{cc}
\scalebox{0.8}{
\scriptsize
\begin{tikzpicture}[>=stealth'] 
 \tikzstyle{lplace}=[circle,thick,draw=black!75,minimum size=4mm]
 \tikzstyle{place}=[circle,thick,draw=black!75,minimum size=5mm]
 \tikzstyle{invplace}=[circle,thick,draw=black!00,minimum size=0mm]
 \tikzstyle{transition}=[rectangle,thick,draw=black!75,minimum size=5mm]
 \tikzstyle{edge}=[->,thick,draw=black!75]
 \tikzstyle{edgered}=[->,thick,draw=red!75]
 \tikzstyle{edgeblu}=[->,thick,draw=blue!75]
 \path ( 1, 1) node [place] (p1) {}
       ( 4, 2) node (a) {}
       ( 4, 0) node (b) {}
       ( 1, 1) node [xshift=0mm,yshift=-5mm] {$(\atom{a},\ast)$}
       ( 1, 1) node [token] {}
       ( 2, 1) node [transition] (c) {\atom{a}}
       ( 5, 1) node [xshift=-10mm,yshift=-4mm] {$(\atom{a},(\setenum{\atom{a}},\atom{a},\lending))$}
       ( 5, 1) node [place] (p5) {}
       ( 2, 1) node [xshift=8mm,yshift=5mm] {$(\setenum{\atom{a}},\atom{a},\lending))$}
       ( 5, 1) node [lplace] {};
       \draw [-,thick]  (c) .. controls +(up:10mm) .. (4,2);
       \draw [edge] (4,2) .. controls +(right:9mm) .. (p5);
       \draw [edge] (p1) to (c);
       \draw [-,thick] (p5) .. controls +(down:10mm) .. (3,0);
       \draw [edge] (3,0) .. controls +(left:9mm) .. (c);
       \path (1,-0.5) node [invplace] {};
\end{tikzpicture}}
& \hspace{15pt}
\scalebox{0.8}{
 \scriptsize\begin{tikzpicture}[>=stealth'] 
 \tikzstyle{lplace}=[circle,thick,draw=black!75,minimum size=4mm]
 \tikzstyle{place}=[circle,thick,draw=black!75,minimum size=5mm]
 \tikzstyle{transition}=[rectangle,thick,draw=black!75,minimum size=5mm]
 \tikzstyle{edge}=[->,thick,draw=black!75]
 \tikzstyle{edgered}=[->,thick,draw=red!75]
 \tikzstyle{edgeblu}=[->,thick,draw=blue!75]
 \path ( -1, 1) node [place] (p1) {}
       ( -1, 1) node [yshift=5mm] {$(\atom{a},\ast)$}
       ( -1, 1) node [token] {}
       ( 5, 2) node [transition] (c) {\atom{c}}
       ( 5, 2) node [xshift=0mm,yshift=4mm] {$t_2$}
       ( 5, 1) node [transition] (b) {\atom{b}}
       ( 5, 1) node [xshift=0mm,yshift=4mm] {$t_3$} 
       ( 1, 1) node [transition] (a) {\atom{a}}
       ( 1, 1) node [xshift=-4mm,yshift=4mm] {$t_1$}
       ( 1, 2) node [place] (p2) {}
       ( 1, 2) node [yshift=5mm] {$(\atom{a},t_1)$}
       ( 1, 0) node [place] (p3) {}
       ( 1, 0) node [xshift=-6mm] {$(\atom{b},t_1)$}
       ( 1, 0) node [lplace] {}
       ( 3, 1) node [place] (p6) {}
       ( 3, 1) node [yshift=4mm,xshift=4mm] {$(\atom{a},t_3)$}
       ( 3, 2) node [place] (p5) {}
       ( 3, 2) node [yshift=5mm] {$(\atom{a},t_2)$}
       ( 7, 2) node [place] (p4) {}
       ( 7, 2) node [xshift=0mm,yshift=-4.5mm] {$(\atom{c},t_1)$}
       ( 6, 3) node [place] (p7) {}
       ( 6, 3) node [xshift=-6mm] {$(\atom{c},t_2)$}
       ( 4, 3) node [place] (p8) {}
       ( 4, 3) node [xshift=6mm] {$(\atom{c},\ast)$}
       ( 7, 3) node [place] (p9) {}
       ( 7, 3) node [xshift=0mm,yshift=-4.5mm] {$(\atom{c},t_3)$}
       %( 7, 3) node [place] (p10) {}
       %( 7, 3) node [xshift=10mm] {$((\atom{a},\atom{b}),\ast)$}
       %( 7, 2) node [place] (p11) {}
       %( 7, 2) node [xshift=10mm] {$((\atom{b},\atom{a}),\ast)$}
       ( 7, 1) node [place] (p12) {}
       ( 7, 1) node [xshift=0mm,yshift=-4.5mm] {$(\atom{b},t_3)$}       
       ( 7, 0) node [place] (p13) {}
       ( 7, 0) node [xshift=-7mm] {$(\atom{b},t_2)$} 
       ( 5, 0) node [place] (p14) {}
       ( 5, 0) node [xshift=-6mm] {$(\atom{b},\ast)$}
%       ( 7, 3) node [token] {}
%       ( 7, 2) node [token] {}
       ( 5, 0) node [token] {}
       ( 4, 3) node [token] {};
       \draw [edge] (p1) to (a);
       \draw [edge] (a) to (p2);
       \draw [edge] (p3) to (a);
       \draw [edge] (a) to (p6);
       \draw [edge] (a) to (p5);
       \draw [edge] (p5) to (c);
       \draw [edge] (p6) to (b);
       \draw [edge] (p14) to (b);
       \draw [edge] (p8) to (c);
%       \draw [edge] (p11) to (b);
%       \draw [edge] (p11) to (c);
%       \draw [edge] (p10) to (b);
%       \draw [edge] (p10) to (c);
       \draw [edge] (c) to (p4);
       \draw [edge] (c) to (p7);
       \draw [edge] (c) to (p9);
       \draw [edge] (b) to (p12);
       \draw [edge] (b) to (p13);
       \draw [edge] (b) to (p3);
 \end{tikzpicture}}
\end{tabular}}
\end{center}
\vspace{-15pt}
\caption{Two contract nets constructed from \pcl contracts.} 
\label{fig:carl-net-compl}
\end{figure}
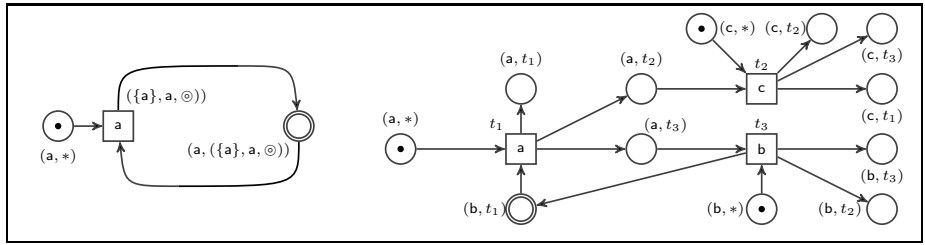

\begin{example}\label{ex:carl-net-compl}
Consider the \pcl contract with formula $\atom{a}\coimp \atom{a}$
(the other components are immaterial for the sake of the example).
The associated LPN is in Fig.~\ref{fig:carl-net-compl}, left.
The transition $(\setenum{\atom{a}},\atom{a},\lending))$, labeled $\atom{a}$, 
can be executed at the initial marking, as the unmarked place in the preset 
is a lending place. 
The reached marking contains no tokens, hence it is honored.
This is coherent with the fact that $\atom{a}\coimp \atom{a} \vdash \atom{a}$
holds in \pcl\!.
\end{example}

\begin{example}\label{ex:a-non-protetto}
Consider the \pcl contract specified by the theory
\[
  \Delta
  \;\; = \;\;
  \setenum{
    \atom{b}\coimp\atom{a},\; \atom{a}\imp\atom{c},\; \atom{a}\imp\atom{b}
  }
\]
The associated LPN is the one on the right depicted in Fig.~\ref{fig:carl-net-compl}.
The transitions are 
$t_1 = (\setenum{\atom{b}},\atom{a},\lending)$, 
$t_2 = (\setenum{\atom{a}},\atom{c},\nolending)$ 
and $t_3 = (\setenum{\atom{a}},\atom{b},\nolending)$. 
Initially only $t_1$ is enabled, 
lending a token from place $(\atom{b},t_1)$. 
This leads to a marking where both $t_2$ and $t_3$ are enabled, 
but only the execution of $t_3$ ends up with an honored marking.
The marking reached after executing all the actions is honored.
This is coherent with the fact that 
$\Delta \vdash \atom{a} \land \atom{b} \land \atom{c}$
holds in \pcl\!.
\end{example}

Since all the transitions consume the token from the places $(\atom{a},\ast)$ 
(where $\atom{a}$ is the  label of the transition), 
and these places cannot be marked again, 
it is easy to see that each transition may occur only once.
Hence, the net associated to a contract is an occurrence net.
If two transitions $t, t'$ have the same label (say $\atom{a}$), then
they cannot belong to the same state of the net. 
In fact, transitions with the same label share the same input place
$(\atom{a},\ast)$. 
This place is not a lending one, and
has no ingoing arcs, hence only one of the transitions with the same
label may happen. 
The notion of correctly labeled net lifts obviously
to contract nets.

\begin{myproposition}\label{prop:pnl-contracts-occurrence}
For all \pcl contracts $\cname$, the net $\contrtopnl(\cname)$ is  correctly labeled.
\end{myproposition}

A relevant property of $\contrtopnl$ is that it 
is an homomorphism with respect to contracts composition. 
Thus, since both $\ccomp$ and $\oplus$ are associative and commutative,
we can construct a physical contract from a set of logical contracts
$\cname_1 \cdots \cname_n$ componentwise, i.e.\ by composing the contract nets
$\contrtopnl(\cname_1) \cdots \contrtopnl(\cname_n)$.

\begin{myproposition}\label{prop:composition homomorphic}
For all $\cname_1, \cname_2$, we have that $\contrtopnl(\cname_1\mid\cname_2) \sim \contrtopnl(\cname_1)\oplus\contrtopnl(\cname_2)$.
\end{myproposition}

In Theorem~\ref{th:pnl conf are states} below we state 
the main result of this section, namely that our construction 
maps the agreement property of \pcl contracts into 
weak termination of the associated contract nets.
To prove Theorem~\ref{th:pnl conf are states}, 
we exploit the fact that $C$ is a set of provable atoms in the logic 
iff $(C,\emptyset)$ is a configuration of the associated contract net.

\begin{mylem}\label{lem:pnl conf are states}
Let $\cname = \seq{\Delta,\aname,\princsym,\payoffsym}$ be a \pcl contract, 
and let $\contrtopnl(\cname) = (O, \aname, \princsym, \Omega)$. 
% be the associated contract net. 
For all $C \subseteq \atoms$, 
$\Delta \vdash \bigwedge C$ iff 
there exists $m\in \mathsf{M}(O)$ such that $\mu(m) = (C,\emptyset)$.
\end{mylem}

\begin{mythm} \label{th:pnl conf are states}
%A \pcl contract 
$\cname$ admits an agreement iff 
$\contrtopnl(\cname)$ weakly terminates in $\payoffsym$.
\end{mythm}

We now specialize Theorem~\ref{th:weakly term and accordance},
which allows for compositional verification of choreographies.
Assuming a choreography specified as a \pcl contract $\cname$,
we can $(i)$ project it into the contracts $\cname_1 \cdots \cname_n$
of its participants, 
$(ii)$ construct the corresponding LPN contracts
$\contrtopnl(\cname_1) \cdots \contrtopnl(\cname_n)$, and 
$(iii)$ individually refine each of them into a service implementation.
If the original choreography admits an agreement, then
the composition of the services weakly terminates,
i.e.\ it is correct w.r.t.\ the choreography.

\begin{mythm}\label{th:lpn-refinement}
Let $\cname = \cname_1\mid \cdots \mid\cname_n$ admit an agreement,
with $\payoffsym[i]$ goals of $\cname_i$.
If $\pname_i$ refines $\contrtopnl(\cname_i)$ 
for $i \in 1..n$,
then $\pname_1 \oplus \cdots \oplus \pname_n$ weakly terminates in 
$\payoffsym[1] \cup \cdots \cup \, \payoffsym[n]$. 
\end{mythm}

The notion of urgency in contract nets correspond to that
in the associated \pcl contracts (Theorem~\ref{th:enc-u}).

\begin{mythm} \label{th:enc-u}
For all \pcl contracts $\cname$, 
and for all $X \subseteq \atoms$,
\(
  \curgent[\cname]{X}
   = 
  \curgent[\contrtopnl(\cname)]{X}  
\).
\end{mythm}

% \begin{example}
% For the \pcl\ contract induced by 
% $\varphi = (a \imp b) \ \land \ (b \coimp a)$,
% we have:
% \begin{align*}
%   \enc[\urgent]{\varphi} \; = \;\;
%   & (\amod[R]{\atom a} \imp \amod[R]{\atom b})
%   \;\land\;
%   (\amod[!]{\atom a} \imp \amod[U]{\atom b})
%   \;\land\;
%   (\amod[R]{\atom b} \coimp \amod[U]{\atom a})
%   \;\land\;
%   \\
%   & (\amod[!]{\atom a} \imp \amod[U]{\atom a})
%   \;\land\;
%   (\amod[!]{\atom b} \imp \amod[U]{\atom b})
%   \;\land\;
%   (\amod[U]{\atom a} \imp \amod[R]{\atom a})
%   \;\land\;
%   (\amod[U]{\atom b} \imp \amod[R]{\atom b})
% \end{align*}
% We have that 
% $\enc[\urgent]{\varphi} \vdash \amod[U]{\atom a}$ and
% $\enc[\urgent]{\varphi} \not\vdash \amod[U]{\atom b}$;
% also, $\enc[\urgent]{\varphi},\amod[!]{\atom a} \vdash \amod[U]{\atom b}$.
% This is coherent with the fact that, in the corresponding contract net 
% $N'' \oplus N'$ in Fig.~\ref{fig:a e b che si implicano a vicenda},
% only {\atom a} is urgent at the initial marking,
% while {\atom b} becomes urgent after {\atom a} has been fired.
% \end{example}

\begin{example}
Recall from Ex.~\ref{ex:pcl:urgent} that,
for $\cname = \seq{\setenum{{\atom a} \imp {\atom b}, {\atom b} \coimp {\atom a}},\ldots}$,
we have:
\[
  \curgent[\cname]{\emptyset} = \setenum{\atom a}
  \qquad
  \curgent[\cname]{\setenum{\atom a}} = \setenum{\atom b} 
  \qquad
  \curgent[\cname]{\setenum{\atom b}} = \setenum{\atom a} 
  \qquad
  \curgent[\cname]{\setenum{{\atom a},{\atom b}}} = \emptyset
\]
This is coherent with the fact that, in the corresponding contract net 
$N'' \oplus N'$ in Fig.~\ref{fig:a e b che si implicano a vicenda},
only {\atom a} is urgent at the initial marking,
while {\atom b} becomes urgent after {\atom a} has been fired.
\end{example}

\section{Related work and conclusions} \label{sec:conclusions}

We have investigated how to compile logical into physical contracts.
The source of the compilation is the Horn fragment of 
Propositional Contract Logic~\cite{BZ10lics}, while the target is 
a contract model based on lending Petri nets (LPNs).
Our compilation preserves agreements (Theorem~\ref{th:pnl conf are states}), 
as well as the possibility of protecting services against misbehavior of 
malevolent services.
LPN contracts can be used to reason  compositionally
about the realization of a choreography (Theorem~\ref{th:lpn-refinement}), 
so extending a result of~\cite{Aalst10cj}.
Furthermore, we have given a logical characterization of those \emph{urgent}
actions which have to be performed in a given state.
This notion, which was only intuitively outlined in~\cite{BZ10lics},
is now made formal through our compilation into LPNs (Theorem~\ref{th:enc-u}).

Contract nets seem a promising model for reasoning on contracts: 
while having a clear relation with \pcl contracts, 
they may inherit as well the whole realm of tools that are already 
available for Petri nets. 

The notion of places with a negative marking is not a new one in the Petri nets community,
though very few papers tackle this notion, as the interpretation 
of \emph{negative} tokens does not match the intuition of Petri nets,
where tokens are generally intended as resources.
In this paper we have used negative tokens to model situations
where actions are in a \emph{circular} dependency, 
like the ones arising in \pcl contracts.
Lending places model the intuition that an action can be performed on a \emph{promise},
and a negative token in a place can be interpreted as the promise made, which must be, sooner or
later, \emph{honored}.
Indeed, the net obtained from a \pcl contract is an occurrence net which may
contain cycles, \emph{e.g.} in the net of Ex.~\ref{ex:a-non-protetto}
the transition $t_1$ depends on $t_3$, which in turn depends on $t_1$ 
(and to execute $t_1$ we required to \emph{lend} a token which is after
supplied by $t_3$).
In \cite{StottsG92} the idea of places with negative marking 
is realized using a new kind of arc, 
called \emph{debit} arcs. 
% This choice does not match with our intuition, and furthermore, 
Under suitable conditions, these
nets are Turing powerful, whereas our contract nets do not add expressiveness
(while for LPNs the issue has to be investigated).
In \cite{MM:LLmodels91} negative tokens arise as the result of certain 
\emph{linear} assumptions. 
The relations with LPNs have to be investigated.

{\small \paragraph{Acknowledgments.}
We thank Philippe Darondeau, Eric Fabre and Roberto Zunino 
for useful discussions and suggestions.
This work has been partially supported by
Aut.\ Reg.\ of Sardinia grants L.R.7/2007 CRP2-120 (TESLA) 
CRP-17285 (TRICS) and P.I.A.\ 2010 (``Social Glue''),
and by MIUR PRIN 2010-11 project ``Security Horizons''.}

\bibliographystyle{abbrv}
\bibliography{biblio}

%\newpage
%\appendix
%\input{gentzen.tex}

\end{document}